\documentclass[a4paper,floatfix,rmp,twocolumn,showkeys,superscriptaddress]{revtex4}
  \usepackage[latin1]{inputenc}
  \usepackage{graphicx}
  \usepackage{mathtools}
  \setcitestyle{numbers,square,comma}
  \usepackage{amsmath}

  \usepackage{url}
  \usepackage{breakurl}
  \usepackage[breaklinks]{hyperref}

\newcommand{\argmin}[1]{\underset{#1}{\operatorname{arg}\,\operatorname{min}}\;}

\begin{document}

  \title{Are Dutch and French languages miscible?}
  \providecommand{\CNB}{Departamento de Biolog\'ia de Sistemas, Centro Nacional de Biotecnolog\'ia (CSIC), C/ Darwin 3, 28049 Madrid, Spain. }
  \providecommand{\GISC}{Grupo Interdisciplinar de Sistemas Complejos (GISC), Madrid, Spain. }
  \providecommand{\USC}{Departamento de F\'isica Aplicada, Universidade de Santiago de Compostela, 15782, Santiago de Compostela, Spain. }
  
  \author{Lu\'is F Seoane}
    \affiliation{\CNB}
    \affiliation{\GISC}

  \author{Jorge Mira}
    \affiliation{\USC} 

  \vspace{0.4 cm}
  \begin{abstract}
    \vspace{0.2 cm}

    We study the stability of the cohabitation of French and Dutch in the Brussels-capital region (Belgium). To this aim, we use available time series of fractions of speakers of monolinguals of both tongues as well as the fractions of bilinguals. The time series span a period from the mid-XIX century until 1947, year of the last accepted linguistic census. During this period, French penetrated the Dutch-vernacular region of Brussels and started a language shift that lasts to this day. The available time series are analyzed with a mathematical model of sociolinguistic dynamics that accounts for cohabitation of languages along bilingualism. Our equations are compatible with long-term coexistence of both languages, or with one tongue taking over and extinguishing the other. A series of model parameters (which we constrain by fitting our equations to the data) determine which long-term trajectory (cohabitation or extinction) would be followed. This allows us to estimate whether the empirical data are compatible with a coexistence of Dutch and French -- in physics terms, whether both tongues are {\em miscible}. Our results tilt towards non-coexistence, or coexistence with a starkly-dominated, minority language. The costs of attempting to sustain such sociolinguistic system are discussed. 

  \end{abstract}

  \keywords{Dutch-French cohabitation, Brussels, bilingualism, language shift}

\maketitle

 \section{Introduction}
    \label{sec:1}

    \begin{figure} 
      \begin{center} 
        \includegraphics[width=\columnwidth]{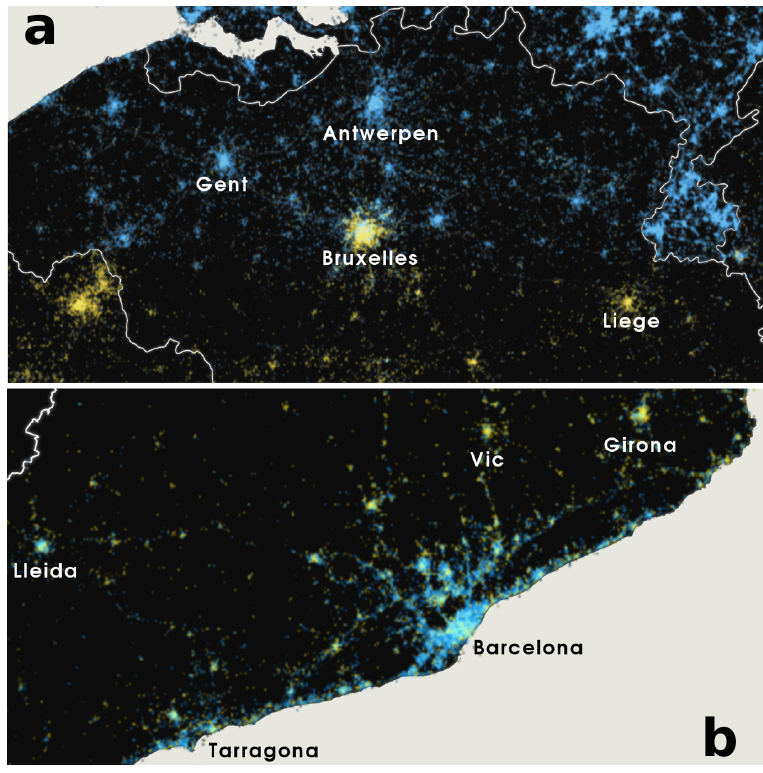}

        \caption{{\bf Miscibility of languages captured by posts in social media.} {\bf a} This map of Belgium shows in blue Twitter posts written in Dutch and in yellow those in French. It is clear the separation of both languages (excepting the area of Brussels), following the border Flanders-Wallonia. {\bf b} Map of Catalonia, where there are also two main cohabitating languages: Catalan and Castillian Spanish. But, in this case, there is a mixing of both (blue: posts in Catalan; yellow: posts in Spanish). Figure modified from \cite{MocanuVespignani2013}.}

        \label{fig:1}
      \end{center}
    \end{figure}

    Languages are one of the human traits that better identify a society. This is clearly seen in some countries with more than one official tongue, where national identities are often linked to linguistic options. Belgium is a remarkable example of this. It is difficult to find a western-world country where a linguistic divide has shaped politics so intensely. Two main tongues dominate Belgium historically: Dutch, in the northern Flanders region; and French, in Wallonia, the southern region. An internal political and administrative boundary was established in 1932 delimiting French- and Dutch-speaking territories \cite{Belxica0}, and setting Brussels aside as a mixed region \cite{Belxica1} (a small German-speaking area also exists on the east side of the country). This boundary, in practice, further consolidated the linguistic divide. The boundary was originally conceived as dynamic: It would be updated according to a linguistic census taken every $15$ years. But in $1961$ several Flemish municipalities refused to abide. The previous boundary (established in $1947$) was adopted as definitive in $1962$ by the Gilson's Act \cite{Belxica2}. That also brought to a halt the periodic censuses and associated sociolinguistic studies -- at least those accepted by both the Flemish and Walloon communities.

    The identity divide can still be observed nowadays in the nomenclature of villages, in signaling across the country, or in geolocalized posts on social media \cite{MocanuVespignani2013, LoufRamasco2021} (Fig.\ \ref{fig:1}{\bf a}). The Brussels-capital region (in French: R\'egion de Bruxelles-Capitale; in Dutch: Brussels Hoofdstedelijk Gewest) stands out because both Dutch and French coexist more closely there -- making it the only officially bilingual French-Dutch region in the country. This puts Brussels forward as a relevant case study for sociolinguistics. Its role as capital city of Belgium (located at Ville de Bruxelles / Stad Brussel) and main center of the European Union further highlights the importance of understanding the sociolinguistic dynamics in this area. 
    
    The clear linguistic boundary seen in the Belgian map contrasts starkly with other cases in which several languages coexist in a same country. Take for example the cohabitation of Catalan and Spanish in Catalonia \cite{SeoaneMira2019}. Despite the acute political conflict (which also involves language and identity), mapping the usage of both tongues across the region reveals a much better mixture (Figure \ref{fig:1}{\bf b}). Looking at the Catalan and Belgian cases through a physics lens, we might think respectively of {\em well-mixed} (as in dissolved salt) versus {\em segregated} phases (as in oil and water). How far away can we take this simile? Might mathematical tools help us model the dynamics of language shift in Belgium, and thus gain some insight on whether both languages are {\em miscible}? Note that Dutch and French are more dissimilar than Catalan and Spanish: may interlinguistic similarity play a role in determining such miscibility? 
    
    In this paper we use a system of ordinary differential equations to study historical data of speakers in different neighborhoods of the Brussels-capital region. By fitting model parameters to the observed sociolinguistic trajectories, we assess the stability of the system of cohabitating languages -- i.e.\ whether both tongues might coexist in the long run, or whether one of them shall go extinct given the model and inferred parameters. We study a historical period that led to the Francization of Brussels \cite{Francization}, in which French displaced Dutch as the most spoken language in this area following a shift in the nation-wide socio-economic balance between southern and northern regions. The available data spans from late-XIX to mid-XX century \cite{datos}. Later socio-economic processes (e.g.\ immigration, the enactment of the internal linguistic boundary, development of regulation to protect either tongue, etc.) have affected the Belgian linguistic scenario in ways that, very likely, supersede the dynamics expected from our fitted model alone. Due to the discontinuation of linguistic censuses, it does not exist comparable data to adapt our model to the newest scenario. However, our aim here is not to forecast how the number of speakers of each tongue might evolve over time, but rather to study the interaction between both languages during a limited period during which we have homogeneously-gathered data. We hope that insights about possible coexistence between both tongues can seed some light on the efforts necessary to keep the linguistic balance in the long term. From a more far-fetched perspective, given the political implications of identity issues, we hope that our results may inform the stability of Belgium as a political entity in a broader sense.

  \section{Methods} 
    \label{sec:2}
    
    \subsection{Data}
      \label{sec:2.2} 
 
      In the Brussels-capital region, speakers of one language are exposed to the other in an effective way. This coexistence fosters sociolinguistic dynamics by which speakers switch or maintain tongues over time, which we attempt to study here. Historical data of such dynamics in the Brussels-capital region are publicly available for a period spanning from the second half of the XIX century to 1947 \cite{datos}. During this period, the so-called Francization of Brussels started, marking a decline of Dutch in favor of French in most neighborhoods of the Brussels-capital region. 

      \begin{figure*} 
        \begin{center} 
          \includegraphics[width=\textwidth]{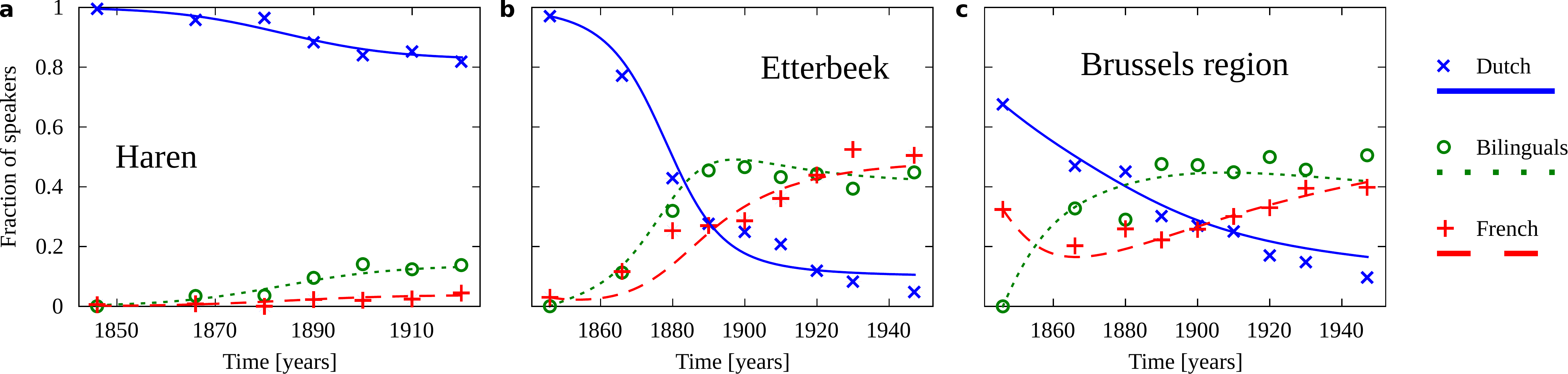}

          \caption{{\bf Examples of fits of data to the model.} The correspondence of lines (fits) and symbols (real data) is indicated at the right. {\bf a} Representation of the best fit among all regions, which corresponds to the municipality of Haren. {\bf b} Representation of the worst fit, which corresponds to the municipality of Etterbeek. {\bf c} Fit to the aggregated data for the whole region of Brussels. }

          \label{fig:2}
        \end{center}
      \end{figure*}
      
      The available censuses show the number of speakers of Dutch, French, and German over time in the 22 municipalities in which the Brussels-capital region was divided during the studied period. These neighborhoods are listed in Tab. \ref{tab:1}. In the time series, we dismissed exclusive German speakers (which were always a minority in every municipality) and normalized the data to obtain fractions of monolingual Dutch and French as well as bilingual speakers. These time series are not homogeneous across municipalities: not all of them have data for the same number of years, as three regions (Haren, Laeken/Laken and Neder-Over-Heembeek) were merged in the Bruxelles-ville / Stad Brussel neighborhood in 1921 and there are no separated data available for them from that year on. We generated an additional time series for the merged Brussels-capital region by aggregating the speakers of each group across all municipalities for each year. We provide the resulting time series in an accessible format. Figure \ref{fig:2} shows this empirical data alongside fits to the model (see next section). 
    
    \subsection{Model and language coexistence} 
      \label{sec:2.1} 

      The central issue in this article is to check the stability of the Dutch-French system in a region where both are simultaneously present. In order to study similar cases we have modified a seminal model of language competition by Abrams and Strogatz \cite{AbramsStrogatz2003} to include scenarios that allow bilingualism \cite{MiraParedes2005}. This framework has been applied to real data from the Spanish Autonomous Communities of Galicia \cite{MiraParedes2005, MiraNieto2011, MusaMira2019} (where Galician and Castillian Spanish are cooficial languages) and Catalonia \cite{SeoaneMira2019} (where Catalan and Castillian Spanish are also cooficial). Furthermore, the stability conditions of the model have been exhaustively worked out, both from the computational \cite{MiraNieto2011} and analytical points of view \cite{OteroMira2013, ColucciOtero2014, ColucciMira2016}.

      The model consists of a system of coupled differential equations that contemplates two monolingual groups ($X$ for Dutch and $Y$ for French in this paper) and a bilingual one, $B$. These groups make up fractions $x$, $y$, and $b$ of speakers respectively within a normalized population ($x + b + y = 1$). The model assumes that the probability that monolingual speakers acquire the opposite language is proportional to a prestige ($s_X$ or $s_Y$) associated to the target tongue. We take $s_X, s_Y \in [0, 1]$ and $s_X + s_Y = 1$; so we can focus on $s \equiv s_X$ (i.e.\ Dutch's prestige). 

      We understand this prestige as in \cite{AbramsStrogatz2003} -- i.e.\, a parameter that reflects, in average, the social or economic opportunities that speakers attribute to the use of each language. More importantly, we take this (and other) parameters as effective properties that emerge from the data, thus constraining effective models that describe each sociolinguistic situation. In other words, the unfolding of the sociolinguistic dynamics is a way to measure the prestige of these tongues and other model parameters. 

      Of all speakers acquiring a new tongue, a fraction $k$ of them retains the old one (hence becoming bilinguals) while $1-k$ of them switch and forget. Since $k$ captures how easy it is to retain both languages, it has been interpreted as a measure of similarity between both tongues (as perceived by their users) \cite{MiraParedes2005, SeoaneMira2017}, and has been termed {\em interlinguistic similarity}. 

      From these assumptions, the probabilities of leaving or entering each group ($X$, $Y$, or $B$) result in a set of differential equations that tells us how the fractions ($x$, $y$, and $b$) of speakers with each linguistic choice evolve over time: 
        \begin{eqnarray}
          {dx \over dt} & = & c\Big[ (b + y)(1-k)s{(1-y)}^a \nonumber \\ 
          && - x\left( (1-k)(1-s){(1-x)}^a + k(1-s){(1-x)}^a \right) \Big], \nonumber \\
          {dy \over dt} & = & c\Big[ (b + x)(1-k)(1-s){(1-x)}^a \nonumber \\ 
          && - y\left( (1-k)s{(1-y)}^a + ks{(1-y)}^a \right) \Big]. 
          \label{eq:01}
        \end{eqnarray}
      Since $b=1-x-y$, a third differential equation is not needed. 

      Many alternative models have been developed over the years to study sociolinguistic contact \cite{BaggsFreedman1990, BaggsFreedman1993, PinascoRomanelli2006, CastelloSan2006, MinettWang2008, KandlerSteele2008, Kandler2009, PatriarcaHeinsalu2009, KandlerSteele2010, CastelloSan2013, ZhangGong2013, HeinsaluLeonard2014, IsernFort2014, ProchazkaVogl2017, MimarGoshal2021}, most of them inspired by the original work by Abrams and Strogatz \cite{AbramsStrogatz2003}. These works include agent-based approaches, mean-field equations drawn from ecological dynamics, or diffusion equations in which the physical distribution of speakers matters. Each model can be suited to different situations and data sets (e.g. reaction-diffusion-like models can be insightful if data is abundantly distributed over space, but sampling is scarce in time \cite{ProchazkaVogl2017}). Several reviews discuss and interpret the existing models and can help navigate the possibilities \cite{CastellanoLoreto2009, SoleFortuny2010, SeoaneMira2017, BoissonneaultVogl2021}. Our data suggests using a mean-field model, of which many in the literature account for bilingualism as well \cite{BaggsFreedman1990, BaggsFreedman1993, HeinsaluLeonard2014, ZhangGong2013}. Our choice of equations allows for stable coexistence solutions (which several models lack) and its stability has been thoroughly studied. 
    
  \section{Results} 
    \label{sec:3}

    \begin{figure*} 
      \begin{center} 
        \includegraphics[width=0.9\textwidth]{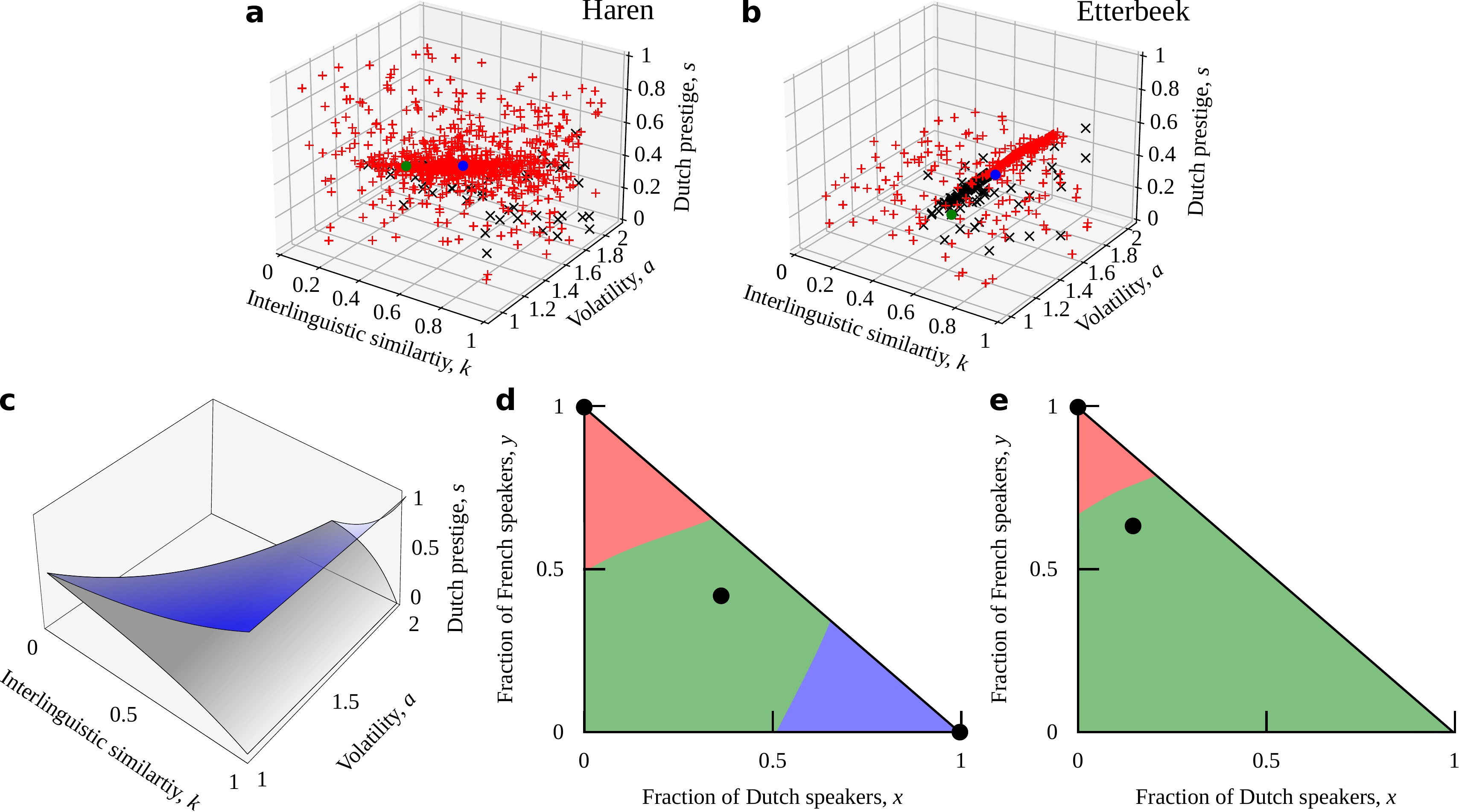}

        \caption{{\bf Empirical data constrains model parameters and possible trajectories in the long term.} {\bf a} Representation of the $\Pi \equiv \{\pi_i\} = \{k_i, a_i, s_i\}$ obtained from the valid fits of the model to the data of Haren (best fit). Red pluses indicate $(k_i, a_i, s_i)$ values that lead to the extinction of one of the languages. Black crosses are compatible with the coexistence of both tongues (depending on initial conditions). The point corresponding to the best fit is marked in green. The weighted average of all valid fits is marked in blue. While the data does not constrain interlinguistic similarity or prestige much, it appears to constrain the expected long-term evolution since most fits lead to the extinction of one of the tongues. {\bf b} Same, for the municipality of Etterbeek (worst fit). The ranges of interlinguistic similarity and prestige are more narrow, but the data seems more ambiguous regarding language extinction. {\bf c} The shadowed surfaces (calculated analytically in \cite{ColucciOtero2014}) define the volume of the $k-a-s$ space for which the model allows long term coexistence -- points outside this region lead to the extinction of one of the languages. Within the volume of coexistence, the eventual fate might depend on the initial conditions. {\bf d-e} Illustration of basins of attraction dependent on initial conditions (as fractions of Dutch monolinguals, $x$; French monolinguals, $y$; and bilinguals, $b=1-x-y$). Qualitatively reconstructed from \cite{MiraNieto2011}. {\bf d} Three attractors (black dots) split up the $x-y$ space depending on whether a unique language would survive (Dutch in the blue basin of attraction, French in the red one), or whether coexistence follows (green). This coexistence attractor observes a symmetric presence of both tongues, with no dominant language. {\bf e} Our model also contains basins of attraction with asymmetric coexistence (green) in which a language (here Dutch) would retain a minoritary presence with respect to the other. }

        \label{fig:3}
      \end{center}
    \end{figure*}

    Following the procedure introduced in \cite{SeoaneMira2019}, we fitted the model described by Eqs.\ \ref{eq:01} to each of our time series separately (including the one with aggregated data from all regions). This method is a fast, stochastic least-square minimization that, for each time series, yields a set $\{a, c, k, s\}$ of parameters that informs us about the sociolinguistic dynamics. To circumvent the stochasticity of the process and the possibility that it might reach local optima, we ran the procedure $1000$ times for each data set. From the resulting collections of plausible parameters we only retained those with meaningful values: $\{a, c, k, s\} >0$ and $0 < \{k, s\} < 1$. Outside these constraints we find unrealistic dynamics (e.g.\ time running backwards or negative fractions of speakers). The number of discarded fits was always negligible (below $5\%$ for any time series), thus in all cases we retained a number $N_F \lesssim 1000$ of successful fits. 

    As a result we obtain a collection of sets of parameters for each empirical time series: $\Pi \equiv \{\pi_i,\>\>i=1, \dots, N_F\}$, such that each $\pi_i \equiv \{a_i, c_i, k_i, s_i\}$ is the result of a single, valid fit. Each $\pi_i$ renders a time evolution of the fraction of speakers ($x(t; \pi_i), y(t; \pi_i)$; Fig.\ \ref{fig:2}), which we compare to the empirical time series to measure the goodness of each set of parameters: 
      \begin{eqnarray}
        \chi^2(\pi_i) &=& {1\over N}\sum_{j=1}^N \left[ \left(x_j - x(t_j; \pi_i)\right)^2 \right. \nonumber \\ 
        && \left. + \left(y_j - y(t_j; \pi_i)\right)^2 \right]^{1/2}. 
        \label{eq:chi}
      \end{eqnarray}
    $N$ is the number of data in the current series. The $x_j$ and $y_j$ are the empirical values labeled by $j=1, \dots, N$. The $t_j$ are the corresponding times (years) at which each empirical value was measured. The $x(t_j; \pi_i)$ and $y(t_j; \pi_i)$ result from evaluating the model with the corresponding parameters $\pi_i$ at the sampling times. Figure \ref{fig:2} shows the fits resulting for the $\pi_i$ with lowest $\chi^2(\pi_i)$ for regions Haren (which is the best fitted region), Etterbeek (worst fitted), and aggregated data for all of Brussels.

    \setlength\tabcolsep{5 pt}
    \begin{table*}[]  
      \begin{tabular}{| l || @{\hspace{1em}}c@{\hspace{1em}} | @{\hspace{1em}}c@{\hspace{1em}} | @{\hspace{1em}}c@{\hspace{1em}} | @{\hspace{1em}}c@{\hspace{1em}} || @{\hspace{1em}}c@{\hspace{1em}} | @{\hspace{1em}}c@{\hspace{1em}} | @{\hspace{1em}}c@{\hspace{1em}} | @{\hspace{1em}}c@{\hspace{1em}} |}
        \hline 
        {\bf Region} & $\hat{a}$ & $\hat{c}$ & $\hat{k}$ & $\hat{s}$ & $\left<{a}\right>$ & $\left<{c}\right>$ & $\left<{k}\right>$ & $\left<{s}\right>$ \\
        \hline 
        1 - Anderlecht & 1.21 & 0.39 & 0.68 & 0.51 & 1.50 & 0.52 & 0.68 & 0.47 \\ 
        2 - Auderghem/Oudergem & 1.44 & 0.37 & 0.67 & 0.42 & 1.56 & 0.66 & 0.69 & 0.42 \\ 
        3 - Berchem-Sainte-Agathe/Sint-Agatha-Berchem & 1.28 & 0.75 & 0.68 & 0.49 & 1.45 & 0.55 & 0.67 & 0.40 \\ 
        4 - Bruxelles-ville / Stad Brussel & 1.92 & 0.46 & 0.69 & 0.47 & 1.75 & 0.50 & 0.69 & 0.45 \\ 
        5 - Etterbeek & 1.11 & 0.30 & 0.67 & 0.39 & 1.52 & 0.64 & 0.64 & 0.41 \\ 
        6 - Evere & 1.46 & 0.13 & 0.74 & 0.23 & 1.49 & 0.43 & 0.66 & 0.40 \\ 
        7 - Forest/Vorst & 1.36 & 0.44 & 0.66 & 0.43 & 1.56 & 0.58 & 0.66 & 0.41 \\ 
        8 - Ganshoren & 1.25 & 0.42 & 0.59 & 0.45 & 1.41 & 0.58 & 0.62 & 0.41 \\ 
        9 - Haren & 0.99 & 1.71 & 0.57 & 0.67 & 1.53 & 0.90 & 0.55 & 0.44 \\ 
        10 - Ixelles/Elsene & 1.33 & 0.09 & 0.65 & 0.33 & 1.61 & 0.38 & 0.60 & 0.44 \\ 
        11 - Jette & 1.40 & 0.31 & 0.65 & 0.47 & 1.54 & 0.46 & 0.67 & 0.43 \\ 
        12 - Koekelberg & 1.08 & 0.07 & 0.75 & 0.14 & 1.50 & 0.55 & 0.70 & 0.44 \\ 
        13 - Laeken/Laken & 1.30 & 0.27 & 0.69 & 0.52 & 1.60 & 0.55 & 0.69 & 0.48 \\ 
        14 - Molenbeek-Saint-Jean/Sint-Jans-Molenbeek & 1.37 & 0.34 & 0.70 & 0.51 & 1.56 & 0.50 & 0.69 & 0.47 \\ 
        15 - Neder-Over-Heembeek & 1.60 & 0.25 & 0.91 & 0.19 & 1.54 & 0.87 & 0.65 & 0.49 \\ 
        16 - Saint-Gilles/Sint-Gillis & 1.12 & 0.26 & 0.64 & 0.39 & 1.55 & 0.47 & 0.67 & 0.42 \\ 
        17 - Saint-Josse-ten-Noode/Sint-Joost-ten-Node & 2.15 & 0.68 & 0.72 & 0.48 & 1.83 & 0.52 & 0.70 & 0.46 \\ 
        18 - Schaerbeek/Schaarbeek & 1.35 & 0.14 & 0.75 & 0.38 & 1.69 & 0.46 & 0.71 & 0.44 \\ 
        19 - Uccle/Ukkel & 1.14 & 0.10 & 0.74 & 0.14 & 1.47 & 0.54 & 0.66 & 0.40 \\ 
        20 - Watermael-Boitsfort/Watermaal-Bosvoorde & 1.11 & 0.30 & 0.66 & 0.42 & 1.54 & 0.55 & 0.66 & 0.40 \\ 
        21 - Woluwe-Saint-Lambert/Sint-Lambrechts-Woluwe & 1.31 & 0.59 & 0.62 & 0.43 & 1.36 & 0.48 & 0.59 & 0.36 \\ 
        22 - Woluwe-Saint-Pierre/Sint-Pieters-Woluwe & 1.21 & 1.01 & 0.58 & 0.46 & 1.27 & 0.54 & 0.61 & 0.39 \\
        23 - {\bf Brussels-capital region} & 1.58 & 0.31 & 0.62 & 0.46 & 1.65 & 0.48 & 0.64 & 0.44 \\ 
        \hline 
      \end{tabular}
      \caption{Numerical outputs of the fits, after processing the sets of values of volatility ($a$), dynamical rate ($c$), interlinguistic similarity ($k$) and Dutch prestige ($s$) for the 19 municipalities of the Brussels-capital region, as well as the 3 that were merged in the Brussels main municipality in 1921. Values $\hat{a}$, $\hat{c}$, $\hat{k}$ and $\hat{s}$ correspond to the best fit for each case (i.e., the fit with the lowest average fit error, $\chi^2$). $\left<{a}\right>$, $\left<{c}\right>$, $\left<{k}\right>$, $\left<{s}\right>$ are the averages weighted across all the valid fits.}
      \label{tab:1}
    \end{table*}    

    Our approach takes advantage of the speed and stochasticity of the fitting procedure to sample the {\em landscape} of model parameters compatible with the observed data. This allows us to visualize how much it is possible to constrain our model dynamics and its parameters given each empirical data series. Figs.\ \ref{fig:3}{\bf a-b} show how the data constrains the parameter landscape for the best (Haren) and worst (Etterbeek) fitted regions (landscapes for other regions are provided in the Supporting Material, but they mostly resemble one of these two extreme cases). Of all combinations of parameters resulting from fits to each data series, we observe that these appear more constrained for Etterbeek (most $\pi_i$ fall within confined values of $k$ and $s$, but not of $a$), while they are more disperse for Haren. 

    In Eqs.\ \ref{eq:01}, some combinations of parameters are compatible with the long-term coexistence of both tongues, while others imply the extinction of one language. Note, first, that this does not depend on the parameter $c$, which is a common factor in Eqs.\ \ref{eq:01} telling us how fast or slow the dynamics unfold. Hence, the parameter landscapes in Figs.\ \ref{fig:3}{\bf a-b} (which show $a$, $k$, and $s$; but not $c$) capture all the information that we can extract from the model regarding long-term language coexistence. Computational \cite{MiraNieto2011} and analytical \cite{OteroMira2013, ColucciOtero2014, ColucciMira2016} studies of the model helped delimit a phase space of expected extinction or cohabitation. The latter is only possible within the surfaces outlined in Fig.\ \ref{fig:3}{\bf c}. Outside this volume, Eqs.\ \ref{eq:01} have only $(x,y)=(1,0)$ or $(0,1)$ as attractors, meaning that one of the two languages would eventually go extinct. Within the coexistence volume, these two attractors might be present together with one of the kind $(x,y) = (x^*, y^*)$ with $0 < x^*,\>y^* < 1$. In these, $x^* + y^* < 1$, thus $b^* > 0$, and bilingualism would always be present alongside monolinguals of either language. Which attractor is reached for $\pi_i$ within the coexistence volume depends on the initial fractions of speakers (i.e.\ the initial conditions, Fig.\ \ref{fig:3}{\bf d-e}). 

    In the landscapes of Fig.\ \ref{fig:3}{\bf a-b}, we plot in red all $\pi_i$ outside the coexistence volume and in black all $\pi_i$ within it. We see that, while the parameters appear more constrained for Etterbeek, they are compatible with diverging outcomes. On the other hand, while the parameter landscape appears less constrained for Haren, a vast majority of $\pi_i$ in this region are not compatible with the long-term coexistence of both tongues. 

    \begin{figure*} 
      \begin{center} 
        \includegraphics[width=0.8\textwidth]{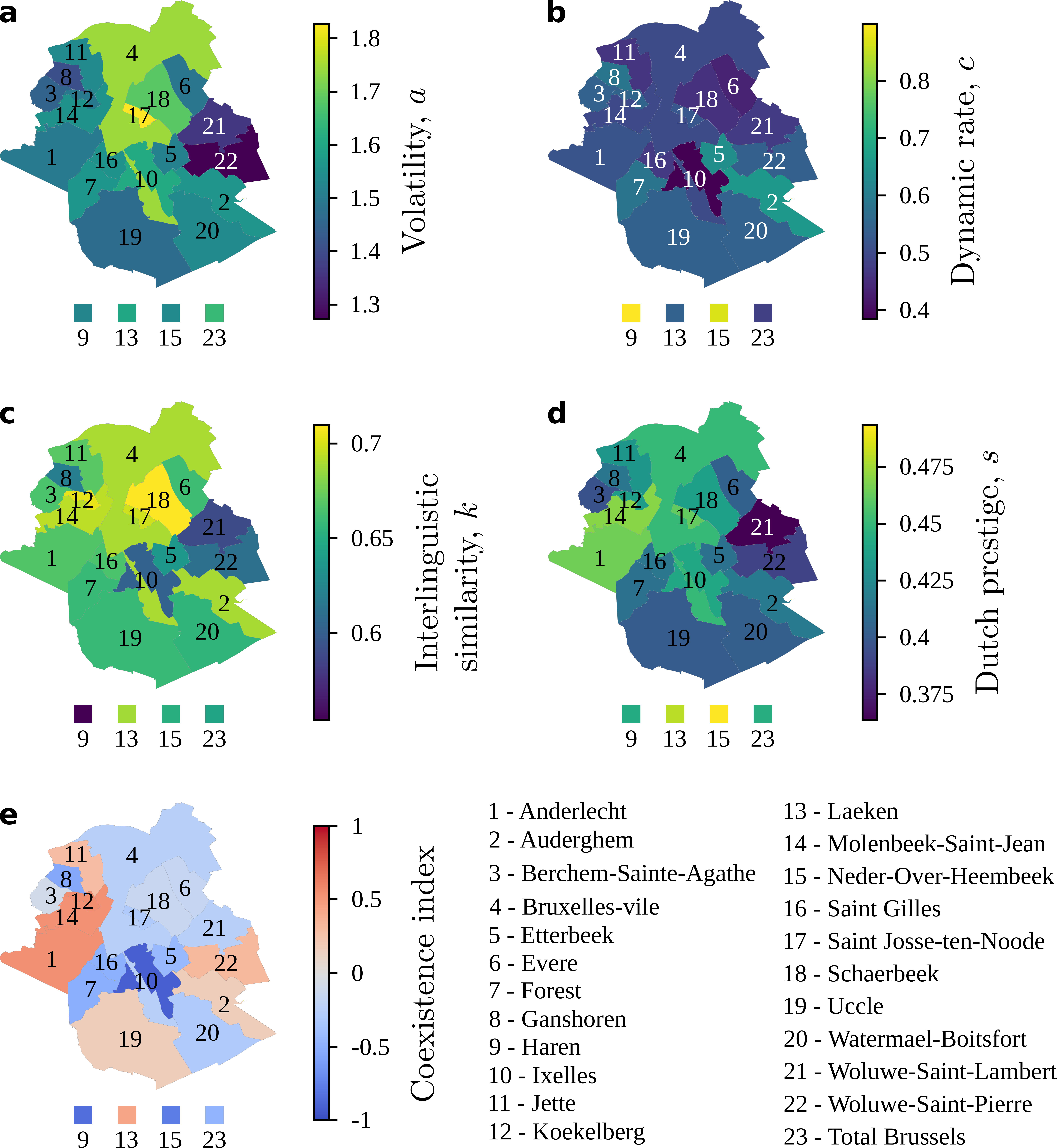}

        \caption{{\bf Geographic distribution of average model parameters and coexistence index.} Maps show the $19$ municipalities with complete time series. Regions $9$, $13$, and $15$ were merged with Bruxelles-ville / Stad Brussel in 1921. Region $23$ stands for the aggregated data for all of Brussels. {\bf a} Average volatility, $\left<a\right>$. {\bf b} Average dynamic rate, $\left<c\right>$. {\bf c} Average interlinguistic similarity, $\left<k\right>$. {\bf d} Average Dutch prestige, $\left<s\right>$. Note that for every municipality it is always lower than that of French. {\bf e} Coexistence index, $\left<g\right>$, for each municipality. }

        \label{fig:4}
      \end{center}
    \end{figure*}

    We summarize the many fits for each data series through the fittest set of parameters: 
      \begin{eqnarray}
        \hat{\pi} &\equiv& \{\hat{a}, \hat{c}, \hat{k}, \hat{s}\} = \argmin{\pi_i}\left\{\chi^2(\pi_i)\right\}. 
        \label{eq:best}
      \end{eqnarray}
    It is also useful to compute an average over valid fits, for which we weight each set of parameters using: 
      \begin{eqnarray}
        \omega_i &\equiv& \exp(-\chi^2(\pi_i)). 
        \label{eq:w}
      \end{eqnarray}
    Thus: 
      \begin{eqnarray}
        \left<a\right> = {1 \over Z}\sum_{i=1}^{N_F} \omega_i a_i, &\text{ with }& Z \equiv \sum_{i=1}^{N_F} \omega_i. 
        \label{eq:aver}
      \end{eqnarray}
    Similar weighted averages are computed for $\left<c\right>$, $\left<k\right>$, and $\left<s\right>$. Both the best and average parameters for each region are reported in Tab.\ \ref{tab:1}, and average parameters are plotted as a map over regions in Fig.\ \ref{fig:4}{\bf a-d} (similar maps showing standard deviation of each parameter are included in the Supporting Material, along with a map of mean $\chi^2$). We observe that the weighted average prestige of Dutch, $\left<s\right>$, is always lower than that of French (Fig.\ \ref{fig:4}{\bf d}); as well as most prestige in the best fits, $\hat{s}$. Interlinguistic similarity is moderately large (average across all regions is $0.656 \pm 0.039$). Note that this does not necessarily capture a similarity of grammar, syntax, or vocabulary alone. Rather, it captures the {\em ease} with which speakers keep both tongues. Strictly-linguistic features might contribute, but this parameter can also be dominated by social factors (e.g. the socioeconomic advantage of communicating across different communities). 

    To summarize the overall trend to extinction or cohabitation of both languages, we introduce a {\em coexistence index}. This quantifies how often fits for a region are compatible with the long-term cohabitation of both languages. We assign $g_i=+1$ if the model with $\pi_i$ evolves a region's initial conditions towards a coexistence attractor, $(x,y) = (x^*, y^*)$; and $g_i = -1$ if the evolution tends to an extinction attractor, $(x,y)=(1,0)$ or $(0,1)$. From \cite{ColucciOtero2014} we have an analytic expression for the coexistence volume. Points outside this volume are assigned $g_i=-1$ trivially. We do not have an analytic expression for the boundaries of attractors as in Fig.\ \ref{fig:3}{\bf d-e}. We evaluated the convergence of these cases numerically (by checking whether they would fall within a radius of $10^{-5}$ around an extinction attractor in the far future). 

    We weight and average the $g_i$ to obtain a coexistence index for each region:
      \begin{eqnarray}
        \left<g\right> = {1 \over Z} \sum_{i=1}^{N_F} \omega_i g_i, 
        \label{eq:04}
      \end{eqnarray}
    as plotted in Fig.\ \ref{fig:4}{\bf e}. Taking the mean of the $\left<g\right>$ across all regions, we find $g_{all} = -0.15 \pm 0.44$: there is an overall tendency towards non-coexistence. The large standard deviation (greater than the mean value) results from averaging over regions with clearly distinct tendencies. Taking separately those that tend to a survival of both languages we get $g_{coex} = 0.38 \pm 0.15$. For regions in which the extinction of one tongue would be more often expected we find $g_{ext} = -0.43 \pm 0.24$. 

    Among coexistence attractors $(x,y) = (x^*, y^*)$, we might find stark asymmetric scenarios in which the presence of one language is greater than the other ($x^* < y^*$ or $y^* < x^*$; Fig.\ \ref{fig:3}{\bf e}). To account for this effect, we computed a modified coexistence index: 
      \begin{eqnarray}
        \tilde{g}_i &=& 2\left[1 - \sqrt{(x^* - y^*)^2}\right] - 1. 
      \end{eqnarray}
    Where we take $(x^*, y^*)\equiv(0,1)$ or $(1,0)$ for extinction attractors. This number is still $-1$ if coexistence is not possible. If long-term cohabitation is possible, $g_i$ is a graded number between $-1$ and $1$ (the closer to $1$, the more symmetric the coexistence). As before, we take, first, the corresponding weighted averages, and then we compute the mean across all regions. We find $\tilde{g}_{all} = -0.34 \pm 0.42$, thus the tendency towards extinction of one of the languages appears strengthened when accounting for asymmetric survival. Separating as before: $\tilde{g}_{coex} = 0.34 \pm 0.12$ and $\tilde{g}_{ext} = -0.52 \pm 0.25$. The fact that this index tilts further towards non-coexistence suggests that most cohabitation scenarios would have a very asymmetric configuration in which a language largely dominates the other.

  \section{Discussion} 
    \label{sec:4}

    In this paper we have studied a period that led to the Francization of Brussels \cite{Francization}, in which French penetrated the formerly Dutch-dominated region of Brussels, eventually substituting the vernacular in most scopes. This shift was fostered by socioeconomic factors that included migration and the international prestige of French, which was also the exclusive administrative language at the time. In consistence with this, our analysis shows a higher average prestige for French across all studied regions. The process (that started in the XIX century) might have been slowed down in the XX century due to the enactment of the linguistic boundary. The official status attained by Dutch or a new shift of economic power to the northern (Dutch-speaking) half of the country might also have contributed to slow down the Francization of Brussels. However, some of the factors favoring French (such as its international prestige and wider usage, probably reinforced by the status of Brussels as center of the European Union \cite{Janssens}) extend their influence to the present day. 

    Our approach is to treat the contact between Dutch and French as a dynamical system limited to a space and time for which we have data series obtained with a consistent methodology, and to observe how these data series constrain a model of sociolinguistic evolution that allows multiple trajectories over time. Some of these trajectories would lead to the coexistence of both languages, while others would imply the extinction of one of the tongues. This would be so if the model would capture all relevant factors that matter in the Dutch-French cohabitation. This last condition is likely not met in the longer run, as effects outside the scope of our study (e.g. large-scale migration as seen in the XX century, further shifts in the socioeconomic balance of power, etc.) come into action. However, we hope that characterizing the dynamics over a limited period of time, with consistently gathered data, we might be able to capture attitudes of speakers towards each-other's tongues. These attitudes would dictate, e.g., perceived prestige or the effort worth making towards sustaining both Dutch and French -- which are the main effects captured by our model. Based on such attitudes, we attempt to infer whether both tongues would likely coexist in the long term. Our insights in the Dutch-French coexistence should be refined in the future if more detailed data, over longer periods of time is gathered, or as complementary models are used to assess, e.g., the influence of the geographic distribution of speakers and how this affects neighboring areas \cite{PatriarcaHeinsalu2009, LoufRamasco2021}. 

    Our results are tilted towards non-coexistence (as indicated by the overall index $g_{all} = -0.15 \pm 0.44$). Where cohabitation appears possible, our results further suggest an asymmetric scenario with a rather minoritary language (as captured by the modified coexistence index $\tilde{g}_{all} = -0.34 \pm 0.42$). Note that the large deviations in both cases arise from tallying together regions with differing outcomes. These results join the observed segregation of both languages in space (Fig.\ \ref{fig:1}{\bf a}) \cite{MocanuVespignani2013, LoufRamasco2021} to suggest, in physics terms, that both languages tend to be {\em not miscible}. Note that we do not imply that a speaker would not be able to learn both tongues -- as it is notably not the case. Rather, our results suggest that the overall situation of French and Dutch (together with the attitudes, risks, and efforts that speakers of each language appear ready to assume regarding the other tongue) are such that natural sociolinguistic dynamics would tend to segregate them. Coexistence is nevertheless {\em not} ruled out, but our results (together with the fact that Brussels is the only officially-bilingual region in Belgium) suggest that the effort needed to enact the cohabitation of Dutch and French is higher than it would be for other coexisting tongues. 

    We take the moderately large interlinguistic similarity (mean value $0.656 \pm 0.039$ across all regions) as a possible indication that these efforts to keep both tongues were being made during the studied period. Note that we are characterizing the contact of a Latin and a Germanic languages, which are much more different than other couples of coexisting tongues (e.g. Catalan and Spanish \cite{SeoaneMira2019}, Fig.\ \ref{fig:1}{\bf b}). This would suggest a smaller interlinguistic similarity for the Dutch-French system (in purely linguistic terms), which should lead to less bilingualism. But the data shows an abundance of this group; hence, socioeconomic factors seem to compensate purely linguistic ones, thus making the use of both tongues more preferable than expected. The stability analysis of our model suggests that, if a balanced cohabitation of both languages is to be obtained, such socioeconomic advantage of bilingualism should be strengthened to move the sociolinguistic system further into the coexistence region. It would pertain to the people of Belgium (as day-to-day users of one or two tongues) and their structures of government to decide whether those efforts are worth making, or whether a tendency to a segregated society is preferred.

\vspace{0.2 cm}

  \section*{Acknowledgments}

    We wish to thank M. Carmen Parafita Couto, from the Leiden University Centre for Linguistics (Leiden University, The Netherlands) for her assistance in this research. Seoane has received funding from the Spanish National Research Council (CSIC) and the Spanish Department for Science and Innovation (MICINN) through a Juan de la Cierva Fellowship (IJC2018-036694-I). Mira is part of iMATUS - USC and of the Program for Consolidation of Research Units of Competitive Reference (GRC), supported by the Xunta de Galicia.

    \begin{figure*} 
      \begin{center} 
        \includegraphics[width=\textwidth]{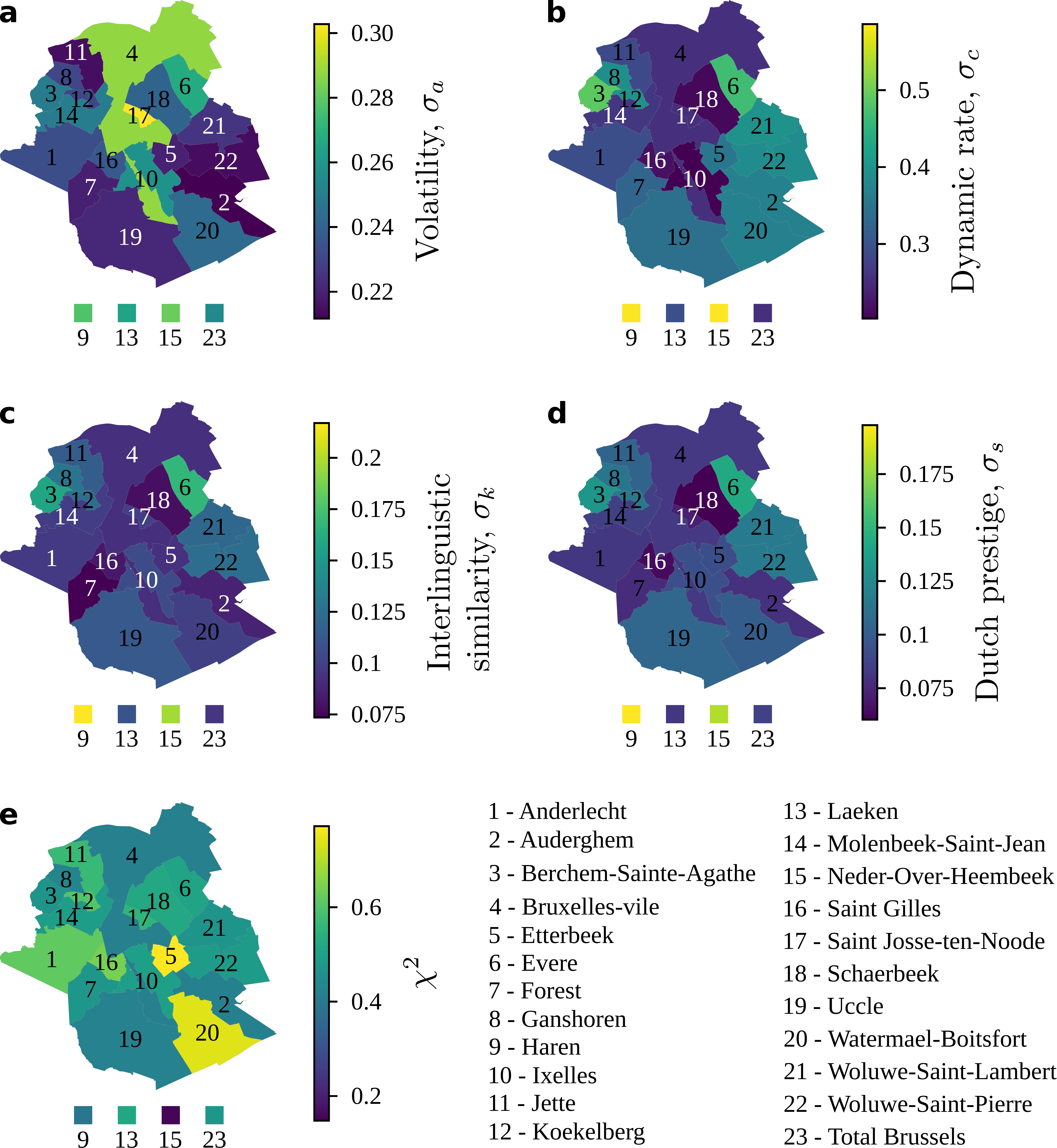}

        \caption{{\bf Standard deviation of parameters and average fit error.} {\bf a} Volatility, $\sigma_a$. {\bf b}
Dynamic rate, $\sigma_c$. {\bf c} Interlinguistic similarity, $\sigma_k$. {\bf d} Dutch prestige, $\sigma_s$. {\bf e}
Average fit error, $\chi^2$. }

        \label{fig:SM1}
      \end{center}
    \end{figure*}

\end{document}